\documentclass[pra,aps,twocolumn,showpacs]{revtex4}
\usepackage{epsfig}
\newcommand{\ket}[1]{|#1\rangle}
\newcommand{\bra}[1]{\langle #1|}

\begin{document}
\title{Measuring Pancharatnam's relative phase for SO(3) 
evolutions using spin polarimetry}
\author{Peter Larsson and Erik Sj\"oqvist\footnote{Electronic 
address: eriks@kvac.uu.se}}
\affiliation{Department of Quantum Chemistry, 
Uppsala University, Box 518, S-751 20 Sweden}
\begin{abstract} 
In polarimetry, a superposition of internal quantal states is exposed
to a single Hamiltonian and information about the evolution of the
quantal states is inferred from projection measurements on the final
superposition. In this framework, we here extend the polarimetric test
of Pancharatnam's relative phase for spin$-\frac{1}{2}$ proposed by
Wagh and Rakhecha [Phys. Lett.  A {\bf 197}, 112 (1995)] to spin
$j\geq 1$ undergoing noncyclic SO(3) evolution.  We demonstrate that
the output intensity for higher spin values is a polynomial function
of the corresponding spin$-\frac{1}{2}$ intensity. We further propose
a general method to extract the noncyclic SO(3) phase and visibility
by rigid translation of two $\pi /2$ spin flippers. Polarimetry on
higher spin states may in practice be done with spin polarized atomic
beams.
\end{abstract}
\maketitle
\section{Introduction}
In polarimetric experiments, a superposition of internal quantal
states evolves in a single spatial beam under a single
Hamiltonian. Information about the Pancharatnam relative phase
\cite{pancharatnam56} is then inferred from projection measurements on
the final superposition. It is thus possible to measure the relative
phase without using spatially separated beams, as in
interferometry. This advantage with polarimetry has proved useful in
measurements of phases of quantal states. Indeed, this technique was
used in the first experiments that measured the cyclic adiabatic Berry
phase \cite{berry84} for two-level systems in terms of polarization of
light \cite{tomita86} and neutron spin
\cite{bitter87,richardson88}. Later, a test of Pancharatnam's relative 
phase in noncyclic spin$-\frac{1}{2}$ polarimetry was put forward
\cite{wagh95} and carried out \cite{wagh00}. In this paper, we 
demonstrate that the polarimetric scheme for spin$-\frac{1}{2}$
proposed in \cite{wagh95} may be extended to spin $j \geq 1$ states 
in noncyclic SO(3) evolution.
 
The polarimetric advantage translates into higher precision when
working with matter waves. This comes from a better utilization of the
particle source in the polarimetric setup, which allows more of the
incoming particles to be used in the experiment \cite{rakhecha01}.
The higher effective intensity improves the precision and enables
experiments with low-flux particle sources \cite{wagh00}. Moreover,
polarimetry is a more robust method, less sensitive to spatial,
mechanical, and thermal disturbances than interferometry. However, the
relative phases can only be measured indirectly in polarimetry, which
complicates the theoretical analysis of the measured data. Here, we
show how this complication can be overcome in SO(3) polarimetry and
propose a general method to extract the relative phase and visibility
in such experiments.
 
Polarimetric tests of noncyclic relative phases for higher spin states
may in practice be done with polarized atomic beams. An interesting
application for such systems could be to verify the noncyclic
geometric phase \cite{samuel88,mukunda93} formula $-m\Omega$ for spin
projections $-j\leq m\leq j$ subtending the geodesically closed solid
angle $\Omega$ in the space of directions in ordinary three
dimensional space. Such an experiment would extend on the atom
interferometry test of the $m$ dependence of the cyclic Berry phase
carried out in Ref. \cite{webb99}.
  
In the following section, Pancharatnam's relative phase is analyzed
for spin$-j$ in SO(3) evolutions. Sec. III describes the noncyclic
relative phase in spin$-\frac{1}{2}$ polarimetry and in Sec. IV it is
extended to $j\geq 1$. The paper ends with the conclusions.
 
\section{Pancharatnam relative phase for SO(3) evolution} 
The Euler representation of SO(3) evolutions may be expressed in terms
of the unitarity ($\hbar = 1$ from now on)
\begin{equation} 
U(\delta,\xi,\zeta) =  
e^{i(\delta + \zeta)J_z} e^{-i2\xi J_y} e^{i(\delta - \zeta)J_z} .  
\end{equation} 
Here, for notational convenience we have expressed $U$ in terms of the
SU(2) parameters $\delta,\xi,\zeta$ \cite{wagh95} that are related to
the standard Euler angles $\alpha,\beta,\gamma$ as 
$\delta = -(\alpha + \gamma)/2, \xi = \beta /2, 
\zeta = -(\alpha - \gamma)/2$. Any $J_z$ eigenket   
$\ket{jm}$ undergoes the SO(3) evolution 
\begin{eqnarray} 
\ket{jm} \rightarrow U(\delta,\xi,\zeta) \ket{jm} 
\end{eqnarray} 
yielding the Pancharatnam relative phase $\Phi_m^{(j)}$ between  
$\ket{jm}$ and $U(\delta,\xi,\zeta) \ket{jm}$ as  
\begin{eqnarray} 
\Phi_m^{(j)} & = & \arg \bra{jm} U(\delta,\xi,\zeta) \ket{jm} 
\nonumber \\ 
 & = & 2m\delta + \arg d^{(j)}_{m,m}(\xi) , 
\end{eqnarray} 
where $d^{(j)}_{m,m}(\xi) \equiv \bra{jm} e^{-i2\xi J_y} \ket{jm}$ is
real valued (see, e.g., Eq. (6.2.16) of Ref. \cite{chaichian98}). This 
latter property implies that $\arg d^{(j)}_{m,m}(\xi)$ only takes the 
values 0 or $\pi$.  Thus, for $m=0$ spin projections these are the only
possible values of Pancharatnam's relative phase in SO(3) evolution.
 
The amount of interference is measured by the visibility ${\cal V}$,  
which in the SO(3) case reads  
\begin{equation} 
{\cal V}_m^{(j)} =  
\big| \bra{jm} U(\delta,\xi,\zeta) \ket{jm} \big| =  
\big| d^{(j)}_{m,m}(\xi) \big| .  
\end{equation}  
For cyclic evolution where $\xi = 0$ (modulo $\pi$), we have maximal
interference contrast ${\cal V}_m^{(j)} = 1$. In the case where 
$m\neq 0$, the angle $\xi = \pi /2$ corresponds to the spin flip 
$m \rightarrow -m$ for which ${\cal V}_m^{(j)}=0$. Depending 
upon the explicit functional form of $d^{(j)}_{m,m}(\xi)$, there 
may exist further $\xi$ values for which the visibility vanishes. 
 
We notice that if $U(\delta,\xi,\zeta)$ is parallel transporting, the
Pancharatnam relative phase can be identified with the noncyclic
geometric phase. In such a case $\Phi_m^{(j)} = -m\Omega$, $\Omega$
being the solid angle enclosed by the path and its shortest geodesic
closure in the space of directions in ordinary three dimensional
space. For example, such a parallel transporting unitarity could be
realized by a sequence of SO(3) rotations along great circles in this
space.

\section{Wagh-Rakhecha setup} 
Consider the Wagh-Rakhecha setup \cite{wagh95} sketched in Fig. 1. A
single beam of spin polarized particles with $j=m=\frac{1}{2}$ and
magnetic moment $\mu$ is sent through a series of devices. A
superposition of the two orthogonal states $\ket{\frac{1}{2},\pm 
\frac{1}{2}}$ is created by rotating $\pi/2$
around an axis perpendicular to the quantization axis of the initial
state. Under the influence of the unitarity $U$ the components of the
superposition acquire opposite Pancharatnam relative phases. Another
$-\pi/2$ rotation is applied and the output intensity is subsequently
measured along the initial quantization axis.

\begin{figure}[htbp]
\centering
\includegraphics[width=8 cm]{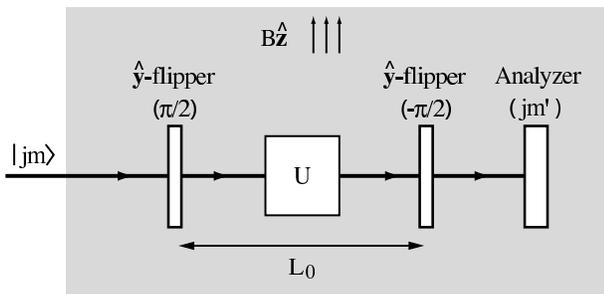}
\caption{Conceptual view of the Wagh-Rakhecha setup for measuring 
noncyclic relative phases in polarimetry. Particles spin polarized 
in the $z$ direction and carrying a magnetic moment $\mu$ are sent
through an SO(3) unitarity, surrounded by two $\pi /2$ spin flippers. 
By rigid translation of the spin flippers at relative distance 
$L_0=n\pi v/|\mu B|$, $n$ integer and $v$ the particle speed, 
the noncyclic relative phase is extracted from the output intensities 
registered at the analyzer.}
\end{figure}

For a cyclic evolution, the output state differs from the initial
state by a rotation of $\Phi_{\frac{1}{2}}^{(\frac{1}{2})}$ about the
initial quantization axis. For a noncyclic spinor evolution, however,
an extra phase shift $\pm \frac{1}{2} \phi$ must be applied to the
spin eigenkets $\ket{u_{\pm}} \equiv \ket{\frac{1}{2},\pm
\frac{1}{2}}$ and the relative phase can thereafter be inferred from
the oscillations of the intensity as measured along the initial
quantization axis, when $\phi$ varies. The extra phase shift $\phi$ is
implemented by a guiding magnetic field $B\hat{{\bf z}}$ put over the
entire setup and the variation of $\phi$ is achieved by 
translating the pair of flippers, keeping their relative distance
$L_0$ fixed. By choosing $L_0=n\pi v/|\mu B|$, $n$ integer and $v$
the particle speed, one obtains the output intensity
\cite{wagh95,remark1}
\begin{equation}
I^{(\frac{1}{2})} = \cos^2 \xi \cos^2\delta + \sin^2 \xi  
\sin^2(\zeta-\phi) .
\label{eq:spinhalf} 
\end{equation} 
This yields the extreme values 
\begin{eqnarray}
I_{\textrm{\small{min}}}^{(\frac{1}{2})} & = & 
\cos^2 \xi \cos^2\delta , 
\nonumber \\ 
I_{\textrm{\small{max}}}^{(\frac{1}{2})} & = & 
\cos^2 \xi \cos^2\delta + \sin^2 \xi ,
\end{eqnarray}
at $\phi = \zeta$ and $\phi = \zeta + \pi /2$, respectively, upon 
translation of the flippers. Now, up to a sign, the Pancharatnam 
relative phase modulo $\pi$ may be obtained as 
\begin{eqnarray}
\cos^2 \Phi_{\frac{1}{2}}^{(\frac{1}{2})} & = & \cos^2 [\delta + 
\arg d_{\frac{1}{2},\frac{1}{2}}^{(\frac{1}{2})} (\xi)] 
\nonumber \\ 
 & = & \cos^2 \delta = \frac{I_{\textrm{\small{min}}}^{(\frac{1}{2})}}
{1-I_{\textrm{\small{max}}}^{(\frac{1}{2})} + 
I_{\textrm{\small{min}}}^{(\frac{1}{2})}} ,
\label{eq:wrpanch}
\end{eqnarray} 
where we have used that 
$\arg d_{\frac{1}{2},\frac{1}{2}}^{(\frac{1}{2})} (\xi)$ is an 
integer multiple of $\pi$. Similarly, we obtain the visibility as 
\begin{equation}
{\cal V}_{\frac{1}{2}}^{(\frac{1}{2})} = \big| \cos \xi \big| = 
\sqrt{1-I_{\textrm{\small{max}}}^{(\frac{1}{2})} + 
I_{\textrm{\small{min}}}^{(\frac{1}{2})}} .  
\label{eq:wrvis}
\end{equation}
Thus, a cyclic evolution is characterized by 
$I_{\textrm{\small{max}}}^{(\frac{1}{2})} =
I_{\textrm{\small{min}}}^{(\frac{1}{2})} = \frac{1}{2}$ 
and spin flip corresponds to the case where
$I_{\textrm{\small{max}}}^{(\frac{1}{2})} = 1$ and
$I_{\textrm{\small{min}}}^{(\frac{1}{2})} = 0$.

The practical advantage of polarimetry may be limited by the modulo
$\pi$ property of the phase measurement, as is clear from the
appearance of $\cos^2\Phi_{\frac{1}{2}}^{(\frac{1}{2})}$ in
Eq. (\ref{eq:wrpanch}). Physically, this arises from the final 
$z$ projection in polarimetry: states with opposite phases give 
the same $| z\rangle$-intensity. Interferometric experiments, on the
other hand, measure modulo $2\pi$, which in particular allows
verification of the Pauli anticommutation \cite{wagh97,wagh99} 
as well as the $\pi$ phase shift associated with the sign of
$d_{\frac{1}{2},\frac{1}{2}}^{(\frac{1}{2})} (\xi)$.

\section{Measuring higher spin phases} 
Consider the Wagh-Rakhecha setup \cite{wagh95} shown in Fig. 1, now
for spin $j\geq 1$ associated with a magnetic moment $\mu$ and
undergoing an arbitrary SO(3) evolution. Prepare a $J_z$ eigenket
$\ket{jm}$ as input and apply a $\pi/2$ flip around the $y$ axis. Each
component of the resulting superposition of $\ket{j\tilde{m}}$ states 
acquires an extra variable phase shift $\tilde{m}\phi$ implemented by the
Zeeman split due to a guiding magnetic field $B\hat{\bf z}$ put over
the entire setup. The SO(3) evolution is followed by a $-\pi/2$ flip
around the $y$ axis, and the output intensity in the $\ket{jm'}$
channel is detected.

This prescription corresponds to the output intensity  
\begin{equation} 
I_{m',m}^{(j)} = 
|\bra{jm'} \widetilde{U} (\delta,\xi,\zeta,\phi) \ket{jm} |^2 
\end{equation} 
with the unitarity 
\begin{equation} 
\widetilde{U}(\delta,\xi,\zeta,\phi) = e^{i\frac{\pi}{2}
J_y}e^{i\phi J_z} U (\delta,\xi,\zeta) e^{-i\phi
J_z}e^{-i\frac{\pi}{2} J_y} .  
\end{equation} 
$I_{m',m}^{(j)}$ may be evaluated by introducing a decomposition of
the $| jm\rangle$ state into spin$-\frac{1}{2}$ states $u_+$ and 
$u_-$ according to (see, e.g., Ref. \cite{chaichian98})  
\begin{eqnarray} 
|jm\rangle & = & \frac{C_{jm}}{(2j)!} \sum_{P} \ket{u_+^{(1)}} \otimes 
\ldots \otimes \ket{u_+^{(j+m)}} 
\nonumber \\ 
 & & \otimes 
\ket{u_-^{(j+m+1)}}\otimes ... \otimes \ket{u_-^{(2j)}}  
\label{eq:sum}
\end{eqnarray} 
with the normalization constant
\begin{equation}
C_{jm} \equiv \sqrt{\frac{(2j)!}{(j+m)!(j-m)!}} , 
\end{equation} 
and by treating the operator $\widetilde{U}$ as a product of 
$\widetilde{U}^{(i)}$ operators, each rotating the spin-$\frac{1}{2}$ 
subspace $i$ separately, as   
\begin{equation} 
\widetilde{U} = \widetilde{U}^{(1)} \otimes \ldots \otimes
\widetilde{U}^{(2j)}. 
\end{equation} 
The summation sign $P$ in Eq. (\ref{eq:sum}) refers to a sum of all
permutations of the labels of the $u$ states. 

For $| jm\rangle$ states, symmetrization brings (2j)! terms in the sum
in Eq. (\ref{eq:sum}). However, it is not necessary to work with a
fully symmetrized state since the symmetrization deals only with the
labeling of the $u_\pm$ states, while the number of $u_+$ and $u_-$
states remains the same. The transformation properties are thus
unaffected by the symmetrization. We only need to consider simplified
states of the form
\begin{equation}
| jm \rangle \equiv C_{jm} u_+^{j+m} u_-^{j-m}.
\label{eq:simply}
\end{equation}
From the spin$-\frac{1}{2}$ case we already know that  
$\widetilde{U}^{(i)}$ acts upon $\ket{u_+^{(i)}}$ and 
$\ket{u_-^{(i)}}$ as 
\begin{eqnarray}
\widetilde{U}^{(i)} \ket{u_+^{(i)}} & = & 
a \ket{u_+^{(i)}} +  b \ket{u_-^{(i)}} , 
\nonumber \\ 
\widetilde{U}^{(i)} \ket{u_-^{(i)}} & = & 
-b^{\ast} \ket{u_+^{(i)}} + a^{\ast} \ket{u_-^{(i)}} ,
\label{eq:ui}
\end{eqnarray}
where
\begin{eqnarray}
a & = & \cos\xi \cos\delta - 
i\sin \xi \sin(\zeta-\phi) , 
\nonumber \\
b & = & i\cos\xi\sin\delta + 
\sin\xi\cos(\zeta-\phi) . 
\end{eqnarray}
Now, $\widetilde{U}$ applied to the state in Eq. (\ref{eq:simply}) 
yields  
\begin{eqnarray}
\widetilde{U}| jm\rangle & = & 
\sum_{m''}| jm''\rangle \frac{C_{jm}}{C_{jm''}} 
\sum_{\nu}{j+m \choose \nu} 
\nonumber \\ 
 & & \times {j-m \choose j+m''-\nu} a^{\nu} b^{j+m-\nu} 
(-b^{\ast})^{j+m''-\nu} \nonumber \\ 
 & & \times (a^{\ast})^{\nu-m-m''} 
 = \sum_{m''}| jm''\rangle \widetilde{U}_{m'',m}^{(j)} , 
\end{eqnarray}
where we have used binomial expansion and the summation range of
the integer $\nu$ is chosen so that the arguments of all factorials 
are positive. Thus, the intensity $I_{m',m}^{(j)}$ for an incident 
$\ket{jm}$ state analyzed in the $\ket{jm'}$ channel reads
\begin{eqnarray}
I^{(j)}_{m',m} & = & 
\left| \widetilde{U}_{m',m}^{(j)} \right|^{2} = 
\left( \frac{C_{jm}}{C_{jm'}} \right)^2 
\nonumber \\ 
 & & \times \Big( \sum_{\nu} (-1)^{\nu}
{j+m \choose \nu} {j-m \choose j+m'-\nu}    
\nonumber \\
 & & \times \big| a \big|^{2\nu - m - m'} 
\big| b \big|^{2j + m + m' - 2\nu} \Big)^2 
\nonumber \\ 
 & = & \frac{(j+m')!(j-m')!}{(j+m)!(j-m)!} 
\nonumber \\ 
 & & \times \Big( \sum_{\nu} (-1)^{\nu} 
{j+m \choose \nu} {j-m \choose j+m' -\nu} 
\nonumber \\ 
 & & \times \Big( I^{(\frac{1}{2})} \Big)^{\nu-\frac{m+m'}{2}} 
\Big( 1- I^{(\frac{1}{2})} \Big)^{j+\frac{m+m'}{2}-\nu} \Big)^2 , 
\nonumber \\ 
\end{eqnarray}
where we have used the identities 
$\big| a \big|^2 = I^{(\frac{1}{2})}$ and $\big| b \big|^2 = 
1 - I^{(\frac{1}{2})}$. Notice that, changing the sign of 
one of $m,m'$ is equivalent to the change of variables 
$I^{(\frac{1}{2})} \rightarrow 1- I^{(\frac{1}{2})}$. Changing 
the sign of both $m$ and $m'$ yields the same intensity due to 
the rotational symmetry of the setup.  

We proceed by looking for extreme points by solving  
\begin{equation}
\frac{\partial I^{(j)}_{m',m}}{\partial\phi} = 
\frac{dI^{(j)}_{m',m}}{dI^{(\frac{1}{2})}} 
\frac{\partial I^{(\frac{1}{2})}}{\partial\phi} = 0.
\end{equation}
From this it is evident that all $I^{(j)}_{m',m}$ have extreme points
at $\phi = \zeta$ or $\phi = \zeta + \pi /2$ corresponding to those in
the spin-$\frac{1}{2}$ case. Thus, the problem of finding $\cos^2
\delta$ and $\big| \cos \xi \big|$ is reduced to using Eqs.
(\ref{eq:wrpanch}) and (\ref{eq:wrvis}) after having determined
$I_{\textrm{\small{min}}}^{(\frac{1}{2})}$ and
$I_{\textrm{\small{max}}}^{(\frac{1}{2})}$ from the measured
intensities.

When $\cos^2 \delta$ has been found we may determine the desired value
of $\cos^2 \Phi_m^{(j)}$ in terms of Chebyshev polynomials as
\begin{eqnarray}
\cos^2 \Phi_m^{(j)} & = & 
\cos^2 \big( 2m\delta + \arg d_{m,m}^{(j)} (\xi) \big)
\nonumber \\ 
 & = & \cos^2 (2m\delta) = [T_{2m}(\cos\delta)]^2 , 
\end{eqnarray}
where we have used that $\arg d_{m,m}^{(j)} (\xi)$ is an integer
multiple of $\pi$. The first few cases are
\begin{eqnarray}
\cos^2 \Phi_{\frac{1}{2}}^{(j\geq \frac{1}{2})} & = & 
\cos^2\delta , 
\nonumber \\
\cos^2 \Phi_{1}^{(j\geq 1)} & = & 
\big( -1 + 2\cos^2\delta \big)^2 , 
\nonumber \\
\cos^2 \Phi_{\frac{3}{2}}^{(j\geq \frac{3}{2})} & = & 
\big( -3\cos\delta + 4\cos^3\delta \big)^2 . 
\end{eqnarray}
Notice here that for all $m=0$, the noncyclic relative phase is 
trivially 0 or $\pi$ independent of $\delta$. Thus, although the 
intensity and $\cos^2\delta$ can be calculated for the $m=0$ cases, 
they are unrelated to the Pancharatnam relative phase because 
$\delta$ does not appear in the phase expression. 

We may also obtain the visibility as a function of 
$\big| \cos \xi \big|$ from the standard expression 
of the matrix elements $d_{m,m}^{(j)}(\xi)$. 
It yields \cite{chaichian98}
\begin{eqnarray} 
{\cal V}_m^{(j)} & = & {\cal V}_{-m}^{(j)} = 
\left| \sum_{\nu} (-1)^{\nu +j+m} 
{j+m \choose \nu} {j-m \choose j+m-\nu} \right. 
\nonumber \\ 
 & & \left. \times \Big( \cos^2 \xi \Big)^{\nu -m} 
\Big( 1- \cos^2 \xi \Big)^{j+m-\nu} \right| .
\end{eqnarray}
The first few cases are
\begin{eqnarray}
{\cal V}_{\frac{1}{2}}^{(\frac{1}{2})} & = & 
\big| \cos \xi \big| , 
\nonumber \\   
{\cal V}_{0}^{(1)} & = & \big| 2 \cos^2 \xi -1 \big| ,
\nonumber \\ 
{\cal V}_{1}^{(1)} & = & \cos^2 \xi ,
\nonumber \\ 
{\cal V}_{\frac{1}{2}}^{(\frac{3}{2})} & = & 
\big| 3 \cos^3 \xi - 2 \cos \xi \big|,
\nonumber \\ 
{\cal V}_{\frac{3}{2}}^{(\frac{3}{2})} & = & 
\big| \cos^3 \xi \big| .  
\end{eqnarray}
Notice here that the visibilities for $m=0$ are well-defined 
in terms of $\big| \cos \xi \big|$.

In general, it is impossible to find closed expressions for $\cos^2
\delta$ in terms of the measured intensities. An interesting
exception, however, is the case of spin-coherent states
\cite{peres95}, characterized by $m=j$. For such states, 
the intensity reads 
\begin{equation}
I_{j,j}^{(j)} = \Big( I^{(\frac{1}{2})} \Big)^{2j} ,
\end{equation} 
which allows a direct evaluation of $\cos^2 \delta$ and $\big| \cos
\xi \big|$ in terms of the extreme values of $I_{j,j}^{(j)}$
according to
\begin{eqnarray} 
\cos^2\delta & = & 
\frac{\sqrt[2j]{I_{j,j;\textrm{\small{min}}}^{(j)}}} 
{1-\sqrt[2j]{I_{j,j;\textrm{\small{max}}}^{(j)}} + 
\sqrt[2j]{I_{j,j;\textrm{\small{min}}}^{(j)}}} ,  
\nonumber \\ 
\big| \cos \xi \big| & = & 
\sqrt{1 - \sqrt[2j]{I_{j,j;\textrm{\small{max}}}^{(j)}} + 
\sqrt[2j]{I_{j,j;\textrm{\small{min}}}^{(j)}}} . 
\end{eqnarray}
The reason for this is that when $m=j$ the state in
Eq. (\ref{eq:sum}) consists of a term with all spin components in 
the same direction, i.e.,  
\begin{equation}
|jj\rangle = 
\ket{u_+^{(1)}} \otimes \ldots \otimes \ket{u_+^{(2j)}} . 
\end{equation}
This implies that the intensity for the $| jj\rangle$ state is nothing
but the product of $2j$ spin-$\frac{1}{2}$ intensities.  

\begin{figure}[htbp]
\begin{center}
\includegraphics[width=8 cm]{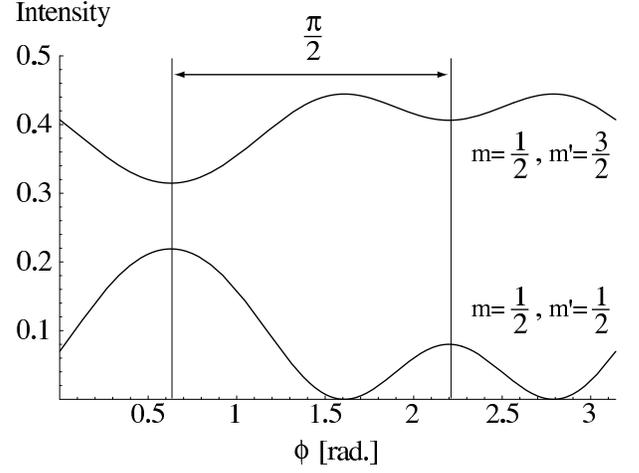}
\end{center}
\caption{Output intensity for $\delta =\zeta =\xi= \pi /5$ as a 
function of the phase shift $\phi$ that is varied upon translation of
the spin flippers. The phase shifts $\zeta$ and $\zeta + \pi /2$ 
are detected by looking for pairs of extreme points at mutual 
distance $\pi /2$.}
\end{figure}

\begin{figure}[htbp]
\begin{center}
\includegraphics[width=8 cm]{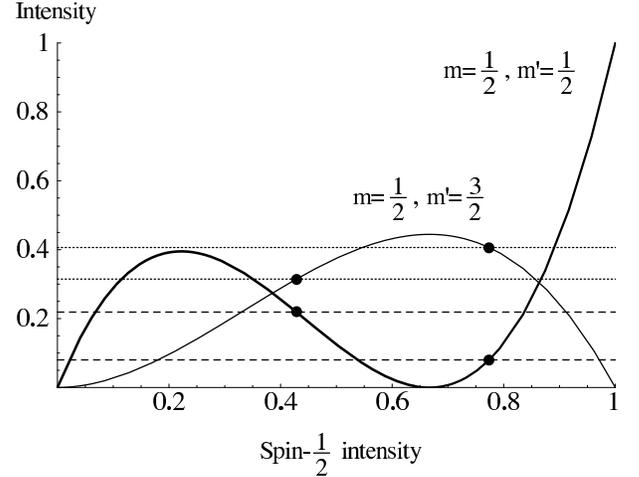}
\end{center}
\caption{Theoretical output intensities in the spin 
projection channels $m' = \frac{1}{2}$ (thick curve) and $m' =
\frac{3}{2}$ (thin curve) for the input state $(j,m) =
(\frac{3}{2},\frac{1}{2})$, as a function of the spin$-\frac{1}{2}$
intensity. The two pairs of horizontal lines correspond to the
extreme values for $m' = \frac{1}{2}$ (dashed line) and 
$m' = \frac{3}{2}$ (dotted line) at phase shifts $\phi = \zeta$ 
and $\phi = \zeta + \pi /2$ shown in Fig. 2. Their 
intersections with the theoretical curves can be matched to 
give the extreme values $I_{\textrm{\small{min}}}^{(\frac{1}{2})} 
\approx 0.43$ and $I_{\textrm{\small{max}}}^{(\frac{1}{2})} 
\approx 0.77$ of the spin-$\frac{1}{2}$ intensity, from which 
the noncyclic relative phase and visibility can be inferred.}
\end{figure}

In the $0< m \neq j$ case, the measured intensity could be a 
nonmonotonous polynomial function of the spin$-\frac{1}{2}$ 
intensity. Thus, there is in general many possible 
$I^{(\frac{1}{2})}$ values for a given measured intensity. 
To remove this ambiguity we may use several intensity  
profiles. To illustrate this point let us consider the 
$j=\frac{3}{2}$ case, where we have 
\begin{eqnarray} 
I_{\frac{3}{2},\frac{1}{2}}^{(\frac{3}{2})} & = & 
3\big( I^{(\frac{1}{2})} \big)^2 \big( 1 - I^{(\frac{1}{2})} \big) , 
\nonumber \\ 
I_{\frac{1}{2},\frac{1}{2}}^{(\frac{3}{2})} & = & 
I^{(\frac{1}{2})} \big( 3I^{(\frac{1}{2})} -2 \big)^2 .  
\label{eq:tc}
\end{eqnarray} 
Both these intensities have extreme values at $\phi=\zeta$ and 
$\phi = \zeta + \pi /2$, as shown in Fig. 2 for $\delta =\zeta = 
\xi = \pi /5$. These extreme values correspond to the horizontal 
lines in Fig. 3 whose intersections with the theoretical curves 
in Eq. (\ref{eq:tc}) can be matched to give the solutions 
$I_{\textrm{\small{min}}}^{(\frac{1}{2})} \approx 0.43$ and
$I_{\textrm{\small{max}}}^{(\frac{1}{2})} \approx 0.77$ from which we
obtain $\cos^2 \delta = \cos^2 \Phi_{\frac{1}{2}}^{(\frac{3}{2})}
\approx 0.65$ and $\big| \cos \xi \big| \approx 0.81$ by using 
Eqs. (\ref{eq:wrpanch}) and (\ref{eq:wrvis}). The latter value may 
be used to compute the visibility as 
${\cal V}_{\frac{1}{2}}^{\frac{3}{1}} = \big| 3\cos^3 \xi - 
2 \cos \xi \big| \approx 0.03$. 

We finally notice that there might be cases where the problem in
assigning unique extreme spin$-\frac{1}{2}$ values cannot be resolved,
because of low visibility and the finite precision in the experimental
data. For example, due to the almost vanishing visibility in the
$(j,m)=(\frac{3}{2},\frac{1}{2})$ case discussed above, there are
crossings near $I^{(\frac{1}{2})} \approx 0.53$ in Fig. 3 that in any
real experiment would potentially be difficult to tell that they in
fact correspond to a spurious solution. This problem may be overcome
either by increasing the resolution of the experiment or by looking at
the intensity in more than two output channels.  It is likely that in
most cases, any ambiguity of this kind can be resolved in this way.

\section{Conclusions}
We have extended the polarimetric setup proposed for
spin$-\frac{1}{2}$ in Ref. \cite{wagh95} and implemented in
Ref. \cite{wagh00} for the same case, to spin $j\geq 1$ in noncyclic
SO(3) evolution. The key feature that makes it possible to extract the
noncyclic relative phase and visibility in such experiments is that
the output intensity for any spin value is a polynomial function of
the corresponding spin$-\frac{1}{2}$ intensity. This entails that the
existence of phase shifts at distance $\pi /2$ corresponding to
extreme intensities is general, and these extreme values can in turn
be used to extract the desired quantities. This procedure becomes
particularly simple in the case of spin-coherent states, where the
noncyclic relative phase and visibility can be expressed directly in
terms of the measured intensities. However, in the general case, such
closed expressions do not exist and measurements in several output
channels is needed.

An apparent extension of the present work is to consider polarimetric
tests of the noncyclic relative phase for spin $j\geq 1$ in SU($2j+1$)
evolution.  Previously, the phase in the SU(3) case has been analyzed
in Ref. \cite{khanna97,arvind97} and a three-channel optical
interferometry experiment of the cyclic SU(3) geometric phase has been
proposed in Ref. \cite{sanders01}. From the perspective of the special
unitary group, it is indeed expected that the Wagh-Rakhecha setup
\cite{wagh95} should be possible to extend to higher spin in SO(3)
evolution since the SU(2) group is locally isomorphic to the group of
rotations in three dimensional space. However, no such isomorphism
exists for higher $j$ and thus there is no simple way to extend the
SU(2) method to higher SU($2j+1$) evolutions using the Wagh-Rakhecha 
setup, because an SU($2j+1 \geq 3$) operator does not work only as a 
rotation operator. 

We hope that the present work will lead to further considerations of
polarimetric phase measurements, in particular in connection to special
unitary transformations on finite dimensional Hilbert spaces, as well
as to experiments that tests the Pancharatnam relative phase using,
e.g., polarized atomic beams.

\section*{Acknowledgment}
The work by E.S. was financed by the Swedish Research Council. 


\begin{thebibliography}{99}
\bibitem{remark1} To be more precise, all the intensities in 
this paper are detection probabilities in the various output 
channels. The measured intensities are proportional to these 
probabilities with the input intensity as proportionality factor.
\end{thebibliography}
\end{document}